\documentclass[11pt]{iopart}
\usepackage{iopams}  
\usepackage{graphicx,xcolor} 
\usepackage[active]{srcltx} 
\usepackage{units}
\usepackage{dsfont}
\usepackage{bbm}

\newcommand{\be}{\begin{eqnarray}}
\newcommand{\ee}{\end{eqnarray}}
\newcommand{\ket} [1] {\vert#1\rangle}
\newcommand{\bra} [1] {\langle#1\vert}
\newcommand{\cfour}{\ket{C_4}}
\newcommand{\uno}{\leavevmode\hbox{\small1\normalsize\kern-.33em1}}
\newcommand{\cluster}{\ket{\Phi^{\mbox{lin}}_4}}
\newcommand{\id}{\mathds{1}}
\newcommand{\rr}{{\mathbbm{R}}}
\newcommand{\LCtilde}{\ket{\widetilde{\rm LC}_6}}

\begin{document}

\title{Optimal verification of entanglement in a photonic cluster state experiment}

\author{H. Wunderlich$^1$, G. Vallone$^{2,3}$, P. Mataloni$^{3,4}$, M. B. Plenio$^1$}

\address{$^1$Institut f\"ur Theoretische Physik, Universit\"at Ulm, Albert Einstein-Allee 11, 89068 Ulm, Germany}
\address{$^2$Centro Studi e Ricerche ``Enrico Fermi'', Via Panisperna 89/A, Compendio del Viminale, Roma 00184, Italy}
\address{$^3$Dipartimento di Fisica della ``Sapienza'' Universit\`{a} di Roma,Roma 00185, Italy}
\address{$^4$Istituto Nazionale di Ottica (INO-CNR), L.go E. Fermi 6, I-50125 Florence, Italy}
%\address{Istituto Nazionale di Ottica (INO-CNR), L.go E. Fermi 6, I-50125 Florence, Italy}
% 

% \ead{custserv@iop.org}
% \eads{\mailto{#1}, \mailto{#2}}

\begin{abstract}

We report on the quantification of entanglement by means of entanglement measures on a four- and a six- qubit cluster state
realized by using photons entangled both in polarization and linear momentum. This paper also addresses the question of the scaling of entanglement bounds from incomplete tomographic information on the density matrix under realistic experimental conditions.
\end{abstract}

\noindent{\it Keywords}: graph state experiment, entanglement measures
%% Uncomment for Submitted to journal title message
%\submitto{\NJP}
\maketitle

\section{Introduction}

Experiments in Quantum Information Science (QIS) rely heavily on multipartite entangled quantum states. Cluster states \cite{BR01}, or more generally graph states, are a particular class of multipartite states that offer a diversity of applications in QIS, ranging from measurement-based quantum computation and error-correction codes to nonlocality tests. Due to the importance of graph states,
a considerable experimental effort has been made to realize them using photons \cite{Walther05, Kiesel05, Lu07, Chen07, Vallone07, cecc09prl} and cold atoms \cite{Mandel03}. Proposals for trapped ions are also pursued \cite{Wunderlich09, Stock09, Ivanov08}.

In this paper we characterize the four- and six-qubit cluster states realized in \cite{Vallone07, cecc09prl,vall10pra} in terms of fidelity, purity and entanglement.
 In particular, we quantify the amount of experimentally generated entanglement using entanglement measures \cite{PV07}.
Cluster states are uniquely defined by a set of correlation operators which generate a group called the stabilizer. An interesting question this paper addresses is how bounds based on measurement results of the generator of the stabilizer alone scale with increasing system size under realistic conditions using the results developed in Refs. \cite{WP09, WP10, WVP10}. We compare the optimal bounds based on such measurements for the fidelity, purity and the robustness of entanglement \cite{Robustness1, Robustness2, Robustness3} as well as the relative entropy of entanglement \cite{relentropy} with the density matrix obtained from all stabilizers in order to answer this question.

\section{Experimental Set-Up}

\begin{figure}[b]
\centering
\includegraphics[width=16cm]{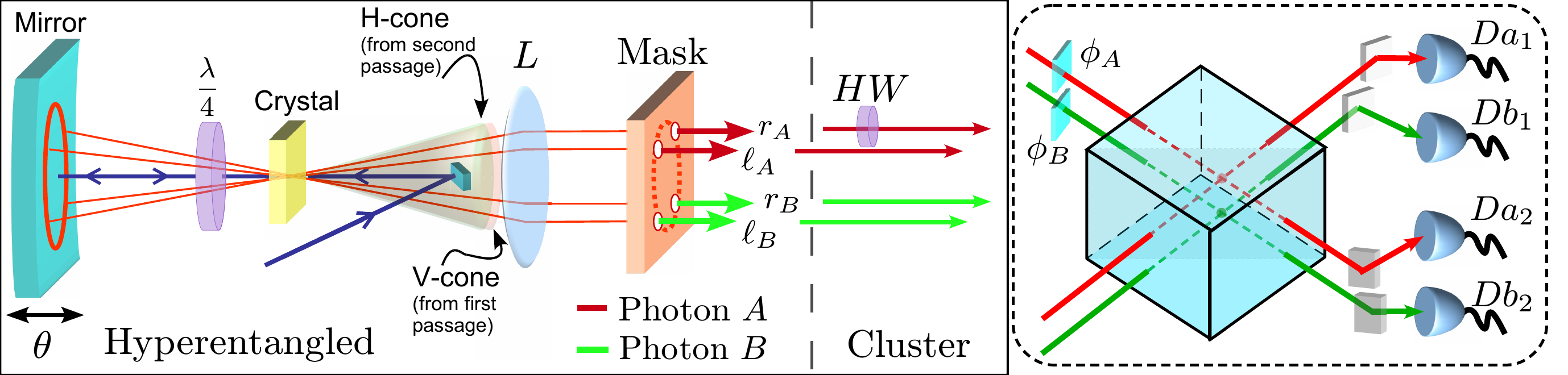}
\caption{left) Source of polarization-path hyperentangled state. right) Measurement setup for the path DOF.}
\label{fig:source}
\end{figure}
Cluster states are particular multiqubit entangled states associated to a graph. In the following we denote the Pauli spin matrices acting on the Hilbert space of qubit $q$ by  $X_q$, $Y_q$, and $Z_q$.
Given a lattice with $n$ vertices and $L$ links, a $n$-qubit cluster state can be defined by associating
a qubit in the superposition state $\ket+=\frac{1}{\sqrt2}(\ket0+\ket1)$ to each vertex and
a control-Z gate CZ$_{ab}=\ket0_a\bra0\otimes\uno_b+\ket1_a\bra1\otimes Z_b$
to each link between vertices $a$ and $b$. 
In an equivalent way, the cluster state is defined as the 
unique eigenvector with positive eigenvalues of the $n$ generators $g_a$ defined as 
$g_a=X_a\prod_{b\in \mathcal N_a}Z_b$, where $\mathcal N_a$ is the set of neighbouring vertices linked with $a$. The set of operators $\{g_a\}$ generate an Abelian group called the stabilizer $\mathcal{S}$ of the underlying graph. Note that eigenstates of the generators with negative eigenvalues $g_a |\{i\}\rangle = (-1)^{i_a} |\{i\}\rangle$ with 
$\ket{\{i\}}=\ket{i_1, ..., i_n}$  (the graph state basis states) 
are  also referred to as graph states in the literature. These states are equivalent to each other
up to single qubit unitary transformations and have therefore the same entanglement properties.

As an example, by considering a graph of four qubits linked in a row, the corresponding cluster is given by
\be
\cluster=\frac{1}{2}(\ket{+00+}+\ket{+01-}+\ket{-10+}-\ket{-11-})\,.
\ee

One way of generating cluster states is using photons. A useful tool to realize multiqubit
states is represented by the so-called hyperentanglement (HE), i.e. the entanglement of two (or more) particles
in several degrees of freedom (DOFs) \cite{kwia97jmo}.
Precisely, by using the source shown in figure \ref{fig:source}, we can generate two photons 
hyperentangled in polarization and path:
\be\label{HE-pi-k}
\ket{\Xi_4}=\frac{1}{\sqrt2}({\ket{H}}_A{\ket{H}}_B+{\ket{V}}_A{\ket{V}}_B)\otimes
\frac{1}{\sqrt2}({\ket{\ell}}_A{\ket{r}}_B-{\ket{r}}_A{\ket{\ell}}_B)\,.
\ee
In the previous equation $\ket H$ ($\ket V$) represent the horizontal (vertical) polarization state and
$\ket r$ and $\ket\ell$ are the two modes (right and left) in which each photon ($A$ and $B$) can be emitted.

The two photons (at degenerate wavelength $\lambda = \unit[728]{nm}$) 
are emitted by the spontaneous 
parametric down conversion (SPDC) process in a
nonlinear type-I $\beta$-barium-borate (BBO) crystal.
The BBO crystal is shined by a vertically polarized continuous wave (cw) Ar$^+$ laser ($\lambda_p = \unit[364]{nm})$.
Polarization entanglement is generated by the double passage (back and forth,
after the reflection on a spherical mirror) of the UV beam.
The backward emission generates the so called V-cone: the SPDC horizontally polarized photons
passing twice through the quarter waveplate (QWP) are transformed into vertical polarized photons.
The forward emission generates the H-cone.
Temporal and spatial superposition are respectively guaranteed by the long coherence time of the UV beam and
by aligning the crystal at a distance from the spherical mirror
which is equal to its radius of curvature.
In this way, the indistinguishability of the two perpendicularly polarized 
SPDC cones creates polarization entanglement:
when two photons are detected it is impossible to know in which pump passage through 
the crystal they have been generated.
It is worth noting that the probability of double pair emission (i.e. four photons) is 
negligible due to the low power of the cw pump beam ($<\unit[100]{mW}$).
By translating the spherical mirror it is possible to change the relative phase between the
states $\ket{HH}_{AB}$ and $\ket{VV}_{AB}$. A lens $L$ located at a focal distance from the crystal
transforms the conical emission into a cylindrical one.

Path entanglement can be generated by exploiting the properties of Type-I phase matching.
The two polarization entangled photons are emitted over two opposite directions of the SPDC cone.
By selecting with a four-holed mask two pairs of correlated modes,
thanks to the spatial coherence property of the source, the two photons are also entangled in path.
We labeled the two pairs of correlated modes, as $r_A-\ell_B$ and $\ell_A-r_B$.
We set the relative phase between the two pair emissions to the value $\varphi = \pi$ by
tilting think glass on the photon paths.
The state expressed in (\ref{HE-pi-k}) encodes 4 qubits into 2 photons \cite{barb05pra,cine05lp}.

\subsection{4-qubit cluster}
The hyperentangled state may be transformed into a cluster state $\cfour$
by using a waveplate with
vertical optical axis and placed on the $\ket r$ mode of the $A$ photon.
The waveplate acts as a $\pi$ phase shift on the state $\ket{V}_B\ket{r}_B$. When applied to $\ket{\Xi_4}$
it generates the following cluster state:
\be
\cfour=
\frac{1}{2}({\ket{H\ell}}_A{\ket{Hr}}_B-{\ket{Hr}}_A{\ket{H\ell}}_B
+{\ket{V\ell}}_A{\ket{Vr}}_B+{\ket{Vr}}_A{\ket{V\ell}}_B)\,.
\ee
By using the correspondence $\ket H_{A,B}\leftrightarrow\ket0_{3,4}$, $\ket V_{A,B}\leftrightarrow\ket1_{3,4}$,
$\ket \ell_{A,B}\leftrightarrow\ket0_{2,1}$, $\ket r_{A,B}\leftrightarrow\ket1_{2,1}$,
the generated state $\ket{C_4}$ is equivalent 
%$\cluster$, $\hs$, 
%$\hsrot$ or $\boxcluster$ 
up to single qubit unitaries to $\cluster$: $\cfour=\mathcal U\cluster= X_1H_1\otimes Z_2\otimes\uno_3\otimes H_4\cluster$, where
$H$ represents the Hadamard gate $H=\frac{1}{\sqrt2}(X+Z)$. The latter relation between $\cfour$ and $\cluster$, 
implies that $\cfour$ is the only common eigenstate of the generators $\tilde g_a=\mathcal Ug_a\mathcal U^{-1}$ obtained
from $g_a$ by changing 
$X_1\rightarrow Z_1$, $Z_1\rightarrow -X_1$, $X_2\rightarrow -X_2$ and $X_4\leftrightarrow Z_4$.

\subsection{6-qubit cluster}
\begin{figure}[b]
\centering
\includegraphics[width=13cm]{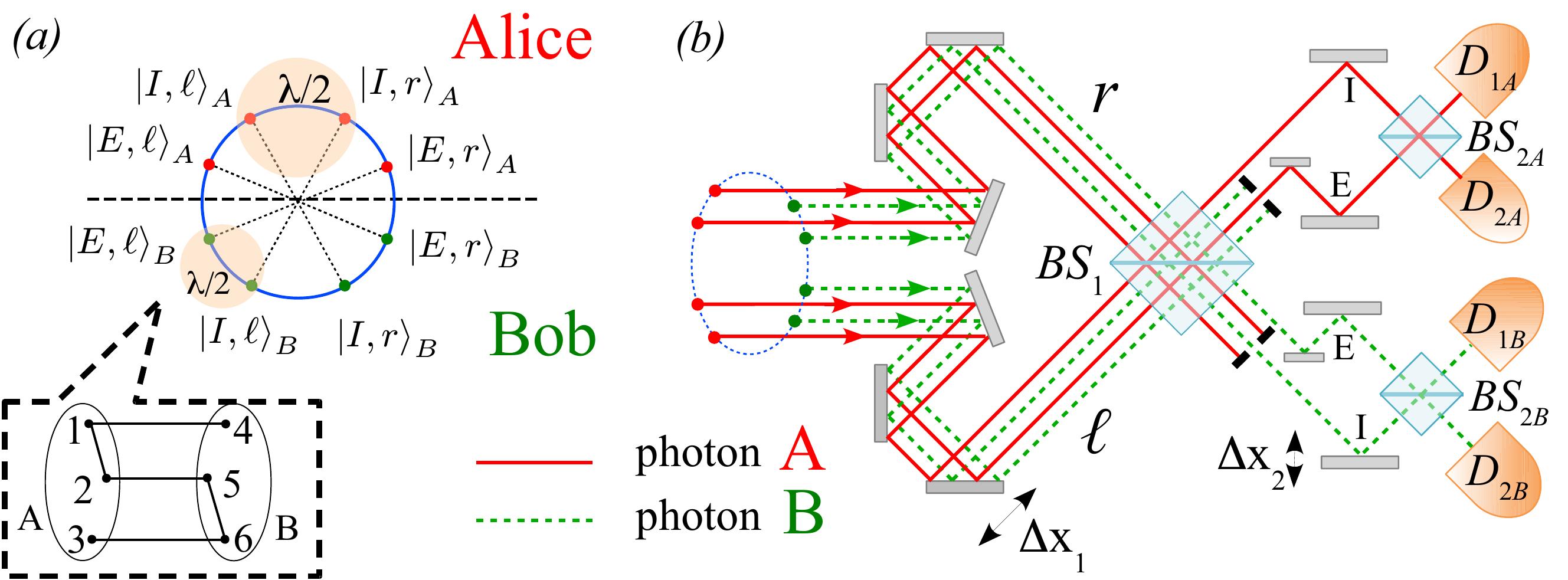}
\caption{Generation and measurement of the 6-qubit linear cluster. 
(a) By selecting 4 pairs of correlated SPDC modes, a 6-qubit polarization-path hyperentangled state can be generated.
Two half waveplates ($\lambda/2$) are used to transform it into a 6-qubit linear cluster state corresponding to the graph
shown in the inset. (b) Two cascade interferometers are used for path measurement. The first $BS_1$ performs the
measurement in the $\{\ket{r},\ket{\ell}\}$ qubit for both photons, 
while the two beam splitters $BS_{2A}$ and $BS_{2B}$ perform the measurement in the $\{\ket{I},\ket{E}\}$ qubit for 
Alice and Bob photon, respectively. Each detection stage $D_i$
is composed of a polarization analyzer
(waveplates and polarizing beam splitter) followed by a single photon detector.
Two translation stages change  the optical delay $\Delta x_{1,2}$ to obtain the correct temporal superposition of the different modes.
}
\label{fig:6qubit}
\end{figure}
It is possible to add more qubits to the state by selecting more optical paths.
Precisely, by selecting four pairs of modes it is possible to generate a
two-photon six-qubit hyperentangled state \cite{vall09pra}. We labeled the four modes on which
each photon can be emitted as $\ket{Er}$, $\ket{E\ell}$, $\ket{Ir}$ and $\ket{I\ell}$,
where $E$ ($I$) stands for external (internal) mode (see figure \ref{fig:6qubit}a)). 
The 6-qubit hyperentangled state can be written as
\newpage
\be\label{HE-6}
\nonumber
\ket{\Xi_6}=\frac{1}{\sqrt2}({\ket{H}}_A{\ket{H}}_B-{\ket{V}}_A{\ket{V}}_B)&\otimes
\frac{1}{\sqrt2}({\ket{r}}_A{\ket{\ell}}_B+{\ket{\ell}}_A{\ket{r}}_B)\otimes
\\
& \otimes \frac{1}{\sqrt2}({\ket{E}}_A{\ket{E}}_B+{\ket{I}}_A{\ket{I}}_B)\,.
\ee
As shown in figure \ref{fig:6qubit}a), two half-waveplates are used to transform the previous state 
into the 6-qubit linear cluster state \cite{cecc09prl, vall10pra}:
\be
\label{lc6}
\LCtilde &= \frac{1}{2} \bigl[\ket{EE}\ket{\phi^{+}}_{\pi}\ket{r
\ell} + \ket{EE}\ket{\phi^{-}}_{\pi}\ket{\ell r} +
\ket{II}\ket{\psi^{+}}_{\pi}\ket{r \ell} -
\ket{II}\ket{\psi^{-}}_{\pi}\ket{\ell r}\bigr]
\ee
where $\ket{\psi^{\pm}}_{\pi}=1/\sqrt{2}(\ket{HV}\pm\ket{VH}$ and $\ket{\phi^{\pm}}_{\pi}=1/\sqrt{2}(\ket{HH}\pm\ket{VV}$. 
$\LCtilde$ corresponds to the graph shown in the inset of figure \ref{fig:6qubit}a) up to single qubit unitaries.
Precisely, $\LCtilde$ is the only common eigenstate (with +1 eigenvalues) of
the generators $\widetilde g_i$ obtained from $g_i$ by changing $X_2 \leftrightarrow Z_2$, $X_3
\rightarrow -Z_3$, $Z_3 \rightarrow X_3$, $X_4 \leftrightarrow
Z_4$ and $X_5 \rightarrow -X_5$.
In order to measure Pauli path operators, two cascade interferometers are implemented (see figure \ref{fig:6qubit}b)).

\section{Results}

\subsection{Quantitative Entanglement Verification}
The detection and quantification of entanglement has become a standard part of quantum information experiments. 
Methods for entanglement detection range from Bell inequalities over entanglement witnesses \cite{GT09} to semidefinite programs \cite{DPS02, NOP09}. In order to quantify entanglement, it is necessary to evaluate an entanglement measure for the state under scrutiny \cite{PV07}. Entanglement measures  have the advantage that they do not only detect entanglement, but they may also provide an operational meaning to the amount of entanglement in a given quantum state. Until today, many entanglement measures have been invented, and the choice of the appropriate measure depends on the specific task \cite{PV07}. 

Here, we choose the global robustness of entanglement \cite{Robustness1, Robustness2, Robustness3} and the relative entropy of entanglement \cite{relentropy}. Both measures are suitable to quantify graph state entanglement for the following reasons: cluster states were introduced as multipartite entangled states that exhibit a particular persistence against noise. While GHZ states become more vulnerable under noise with increasing system size, this is not the case for cluster states \cite{BR01}. Hence, the robustness of entanglement is a measure that quantifies this property. The relative entropy provides an operational meaning for cluster states in the sense that it `counts' the number of entangling gates. As shown in Ref. \cite{MMV07, Anders07}  the relative entropy of entanglement for cluster states is proportional to the number of applied controlled-phase gates.

The relative entropy of entanglement is defined  as \cite{relentropy}
\begin{equation}
E_R (\rho) = \min_{\sigma \in Sep} S(\rho | \sigma),
\end{equation}
where $Sep$ denotes the set of fully separable states, and $S(\rho | \sigma)  = tr[\rho (\log_2 \rho- \log_2 \sigma)]$.

The global robustness of entanglement measures how much noise must be mixed in to a given quantum state such that the mixture becomes separable \cite{Robustness1, Robustness2, Robustness3}:
\begin{equation}
\label{robustness}
R_G(\rho) = \min_{\sigma \in \mathcal{D}, s\in \rr} \{ s:\frac{\rho + s \sigma}{1+s} \in Sep \},
\end{equation}
where $\mathcal{D}$ is the entire Hilbert space.

From a mathematical point of view it is more convenient to relax the global robustness by replacing the set of fully separable states by teh set of PPT states, thus obtaining the following semidefinite program:
\begin{eqnarray}
\label{RPPT}
R_G^{PPT} (\rho) = & \min tr\{\sigma\} &
\\
\mathrm{subject~to}~ & \sigma & \ge 0,
\\
& (\rho + \sigma)^\Gamma & \ge 0 .
\end{eqnarray}
Here $\Gamma$ denotes partial transpostion with respect to a partition of choice. In principal, one could check all possible partitions. In this way we have relaxed the global robustness to a PPT version that can be  formulated as a semidefinite program. Hence, numerical tools such as convex optimization solvers are instantly available to evaluate this measure \cite{sedumi}.
 
\subsection{Four-qubit cluster state}
\begin{figure}
\centering
\includegraphics[width=12cm]{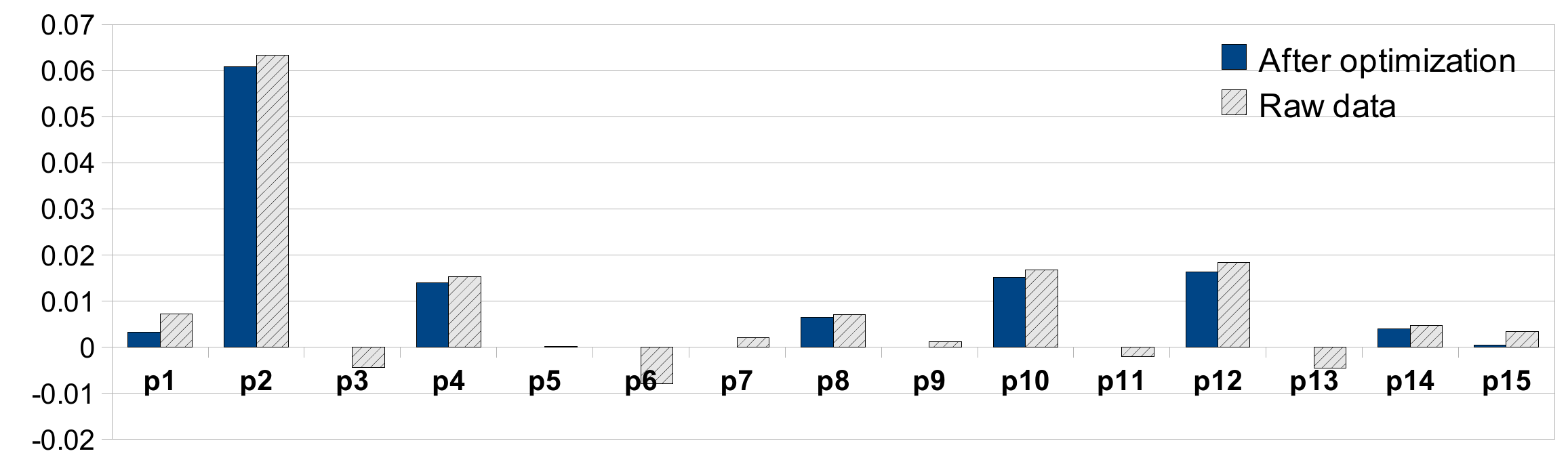}
\caption{Raw and optimized values of $p_{\{k\}}$ for the 4-qubit cluster state. 
Each value corresponds to a given $p_{\{k\}}$ according to the correspondence
$\{k\}\leftrightarrow8k_4+4{k_3}+2{k_2}+{k_1}$.
}
\label{fig:data}
\end{figure}
To verify the creation of the four-qubit cluster state all elements of the stabilizer group were measured \cite{vall08pra}. 
As the measurements were local measurements, also statistics of single Pauli operators are available. These local measurements do not contribute to the fidelity, but they allow us to improve bounds on entanglement measures as those are restricted minimizations that can only improve when more constraints are added \cite{IncompleteTomo1}. Using the measured data
we calculated the raw fidelity $F_{\ket{C_4}} = \frac{1}{16}\sum^1_{\{k_a\}=0}
\langle S_{k_1k_2k_3k_4}\rangle=0.880\pm0.013$ \cite{vall08pra},
where $S_{k_1k_2k_3k_4}=\prod^4_{a=1} (g_a)^{k_a}$ are the 16 stabilizers built as all possible products of generators.
The raw purity is found to be $P (\rho) = tr(\rho^2) =  0.779 \pm 0.005$. 
From the raw data it is possible to obtain the fidelity with all possible graph state bases $\ket{\{i\}}\bra{\{i\}}$
since $\ket{\{i\}}\bra{\{i\}}=\frac{1}{16}\sum_{\{k\}}S_{\{k\}}(-1)^{\bf i\cdot k}$ 
with ${\bf i\cdot k}=\sum^4_{a=1} i_ak_a$. Some of the raw fidelities are negative 
because of experimental inaccuracies and statistical fluctuations of coincidence counts
(the same problem arises in quantum state tomography \cite{jame01pra}). To solve the problem we applied 
a maximum likelihood estimation. We determine the physical density matrix diagonal in the graph state basis and written as
\begin{equation}
\rho_{phys}=\sum_{\{k\}}p_{\{k\}}\ket{\{k\}}\bra{\{k\}}\,,\qquad p_{\{k\}}\geq 0
\end{equation}
that is most compatible with the experimental data. 
The value of the optimized $p_{0000}$ corresponds to the fidelity with the cluster state, 
with value equal to 0.880 which is completely compatible with the raw fidelity.
The other values of the optimized $p_{\{k\}}$ are shown in figure 
\ref{fig:data}. The optimized purity is given by $P_{opt}=\sum_{\{k\}}p^2_{\{k\}}=0.778$ again compatible with the raw value.

While for few-qubit systems the determination of the whole stabilizer is feasible, this is not the case for large graph states. Therefore, it is natural to ask which bounds on the fidelity, purity and entanglement can be obtained from incomplete information on the density matrix \cite{IncompleteTomo1, IncompleteTomo2, IncompleteTomo3}, e.g., from measuring generators of a graph state only; and how do such bounds scale with system size under realistic conditions? 

Let us consider the estimation of the fidelity from information on the generators only. This is formulated as a worst-case estimation \cite{WP09}
\be
F_{min} = \min_{\rho} \{ F(\rho): tr(\rho g_i) = a_i, \rho \ge 0\},
\ee
where $g_i$ are the generators of the graph for $i=1,..,4$ with corresponding measurement outcomes $a_i$, and $g_0 = \id$. Remarkably, this problem can be solved optimally, leading to a solution of the analytic form \cite{WP09}
\be
F_{min} = \max \{0, \frac{\sum_{i=1}^n |a_i|-n+2}{2} \}
\ee
for $n$ qubits and holds for all stabilizer operators with spectrum $\{+1,-1\}$. One may quickly check that the optimal lower bound on the fidelity consistent with the measurements of the generators (see Tab. (\ref{result4})) is given by $F_{min} = 0.846 \pm 0.009$. The relative loss of information on the fidelity is therefore only around 5\%, 
even though only four out of sixteen elements of the stabilizer were determined. 
It is also possible to optimally estimate the purity using only generator measurements. Following the techniques of Ref. \cite{WP10} we obtain a minimal purity consistent with such measurements of $P_{min} = 0.715 \pm 0.014$.

\begin{table}
\caption{Four-qubit cluster state: measurement results of generators}
\label{result4}
\footnotesize\rm
\begin{tabular*}{\textwidth}{@{}l*{15}{@{\extracolsep{0pt plus12pt}}l}}
\br
Generator & Measurement Outcome \\
\mr
$g_1 =- Z_1 \otimes Z_2 \otimes \id_3\otimes\id_4$ & $0.994 \pm 0.001$ \\
$g_2 =- X_1 \otimes X_2 \otimes Z_3 \otimes\id_4$ &  $0.849 \pm 0.003$\\
$g_3 =  \id_1 \otimes Z_2 \otimes X_3 \otimes X_4$ &  $0.937 \pm 0.003$ \\
$g_4 =  \id_1 \otimes \id_2 \otimes Z_3 \otimes Z_4 $ &   $0.911 \pm 0.002$  \\
\br
\end{tabular*}
\end{table}

Quantifying the experimentally created entanglement is achieved by evaluating the global robustness of entanglement and the relative entropy. The density matrix reconstructed from the stabilizer measurements and local observables obtained as a side product of the stabilizer measurements serves as the input for the semidefinite program (\ref{RPPT}), where we evaluate the global robustness with the constraint of positivity of the partial transpose with respect to all partitions. 
We find that the PPT-Robustness is given by 
%\begin{equation}
$R_G^{PPT}  =  2.519 \pm 0.012$. Note that this value represents a lower bound to the global robustness in its standard version (\ref{robustness}).
%\end{equation}
It is straightforward to compute the logartihmic global robustness:
%\begin{equation}
$LR_G^{PPT} =  1.817 \pm 0.005$, which is not far from its desired value of $2$.
%\end{equation}

As in the case of the fidelity one might ask which bound on the entanglement can be obtained from generator measurements alone. Here the estimation is analogously formulated as a minimization of the measure over states consistent with the measurement data
\be
R_{G_{min}} = \min_{\rho} \{ R_G(\rho) : tr(\rho g_i) = a_i, \rho \ge 0\}.
\ee
A lower bound to this problem was derived in Ref. \cite{WVP10}, namely $R_{G_{min}}  = \max\{0, 2^{|B|} (\frac{\sum_{i=1}^n |a_i|-n+2}{2})-1\}$. Here $B$ denotes the smaller set of qubits resulting from a coloring of the system into two colors, say Amber $A$ and Blue $B$ with $|A| \ge |B|$ (see Ref. \cite{MMV07} for more details). With this formula, we attain the following analytic bound on the global robustness based on the outcomes of the generators only \cite{WVP10}:
\begin{equation}
R_{G_{min}}  =  2.384 \pm 0.036 .
\end{equation}
In turn, one can then easily compute an analytic bound on the logarithmic global robustness:
%\begin{equation}
$LR_{G_{min}}  =  1.759$.
%\end{equation}
   
The relative entropy can also be bounded from below using techniques presented in Ref. \cite{WVP10}.
The problem reads:
\be
E_{R_{min}} = \min_{\rho} \{ E_R(\rho) : tr(\rho g_i) = a_i, \rho \ge 0\}.
\ee
A lower bound to this minimization is given by 
\be
\label{bound_relent}
E_{R_{min}} = \max \{0,|B|-\sum_i H(p_i)\},
\ee
where $p_i = \frac{1+a_i}{2}$ and $H(x) = - x \log(x) - (1-x) \log (1-x)$ is the classical entropy function. Then, by merit of Eq. (\ref{bound_relent}) we achieve the following bound on the relative entropy of entanglement:
%\begin{equation}
$ E_{R_{min}} = 1.120 \pm 0.021$. 
%\end{equation}
Using all stabilizer measurements we find a lower bound to the relative entropy of $1.449 \pm 0.013$. Hence, the relative difference of the entanglement bounds of the relative entropy is considerably larger than the relative difference of the estimate of the robustness to its real value.    

\subsection{Six-qubit cluster state}
The six-qubit cluster state is verified utilizing the same techniques as in the four-qubit case. All 64 stabilizer operators were measured and mapped to a density matrix via maximum likelihood, giving a fidelity of
%\begin{equation}
$F= 0.645 \pm 0.006$.  
%\end{equation}
Estimating the fidelity from the generators alone gives
%\begin{equation}
$F_{min} = 0.545 \pm 0.027$.
%\end{equation}

Next, we obtain for the purity
%Purity (all stabilizers):
%\begin{equation}
$P (\rho)  = tr(\rho^2) = 0.424 \pm 0.010$, and the worst-case purity estimate from the generators
%\end{equation}
%\begin{equation}
  $P_{min} = 0.297 \pm 0.015$.
%\end{equation}
We compute the PPT-Robustness for the reconstructed state and find
%\begin{equation}
$R_G^{PPT} =  4.507 \pm 0.047$
%\end{equation}
resulting in a logarithmic PPT-Robustness of
%log robustness:
%\begin{equation}
$LR_G^{PPT} =  2.461$.
%\end{equation}
If one only measures the generators of the stabilizer, one obtains $R_{G_{min}} =  3.360 \pm 0.216$ and $LR_{G_{min}} = \log (1+ R_{G_{min}}) = 2.124$ respectively. This means that despite the lower fidelity and lower fidelity estimate one obtains a higher bound on the entanglement, even though only 6 out of the 64 elements of the stabilizer are used to obtain the bound.

To obtain a bound on a second measure, the relative entropy of entanglement, we use Eq. (\ref{bound_relent} ) to obtain $E_{R_{min}}=1.013 \pm 0.046$. Using all stabilizer measurements, one obtains a lower bound to the relative entropy of $1.492 \pm 0.027$.

\begin{table}
\caption{Six-qubit cluster state: measurement results of generators}
\label{result6}
\footnotesize\rm
\begin{tabular*}{\textwidth}{@{}l*{15}{@{\extracolsep{0pt plus12pt}}l}r}
\br
Generator & Measurement Outcome \\
\mr
$g_1 = X_1 \otimes X_2 \otimes \id_3 \otimes X_4 \otimes\id_5\otimes\id_6 $ & $0.593 \pm 0.008$ \\
$g_2 = Z_1 \otimes Z_2 \otimes \id_3\otimes\id_4\otimes Z_5 \otimes\id_6$  & $0.879 \pm 0.005$ \\
$g_3  =-\id_1 \otimes \id_2\otimes Z_3 \otimes \id_4 \otimes\id_5 \otimes Z_6 $  & $0.998 \pm 0.001$ \\
$g_4 = Z_1 \otimes \id_2 \otimes \id_3  \otimes Z_4 \otimes \id_5 \otimes \id_6$  & $0.997 \pm 0.001$ \\
$g_5 =-\id_1 \otimes X_2 \otimes \id_3 \otimes \id_4 \otimes X_5 \otimes Z_6 $   & $0.791 \pm 0.006$ \\
$g_6 = \id_1 \otimes \id_2 \otimes X_3 \otimes\id_4\otimes Z_5 \otimes X_6 $  & $0.831 \pm 0.006$ \\
\br
\end{tabular*}
\end{table}

\section{Conclusion}
We have presented the creation of four- and six-qubit cluster states using photons. The cluster state entanglement was encoded in path and polarization DOF, thus rendering the state hyperentangled. A summary of the relevant characteristics of the created states is given in Tab. \ref{resultsummary}. The created state could serve as the basis for one-way quantum computation and represents an important step in realizing optical quantum computing. 

The experimentally created entanglement was quantified in terms of entanglement measures, namely the global robustness of entanglement and the relative entropy of entanglement. Our results also give insight into the question how analytic bounds from incomplete tomographic information scale with the system size under realistic noisy conditions. Our results demonstrate that despite the decreasing fidelity and purity of the state, one can still infer higher amounts of entanglement with a number of observables which is linear in the number of constituents. 

\begin{table}[b]
\caption{Summary of experimental results}
\label{resultsummary}
\footnotesize\rm
\begin{tabular*}{\textwidth}{@{}l*{15}{@{\extracolsep{0pt plus12pt}}l}}
\br
   & Fidelity & Purity & Minimal Global Robustness  & Minimal Relative Entropy \\
\mr
4 qubits & $0.880 \pm 0.006$ & $0.778 \pm 0.005$ & $2.519 \pm 0.012$ & $1.449 \pm 0.013$\\
6 qubits & $0.645 \pm 0.006$ & $0.424 \pm 0.010$ & $4.507 \pm 0.047$ & $1.492 \pm 0.027$\\
\br
\end{tabular*}
\end{table}

\section*{Acknowledgments}
We thank E. Pomarico, R. Ceccarelli and G. Donati for their contribution 
in the measurements presented in \cite{Vallone07, vall08pra, cecc09prl, vall10pra}.

This work was supported by the EU Integrated Project Q-ESSENCE, the EU STREP project HIP and by an Alexander von Humboldt Professorship.

 \section*{References} 
% \bibliographystyle{unsrt}% Produces the bibliography via BibTeX.
% \bibliography{../../../bibliografia/qi-bibliografia}% Produces the bibliography via BibTeX

\end{document}